\documentclass[12pt]{iopart}
\usepackage{graphicx}
\usepackage{iopams}
\usepackage[pdftex,dvipsnames,usenames]{color}

\def \UN{{\emptyset}}

\begin{document}

\title{The entropic cost to tie a knot}

\author{M Baiesi$^1$, E Orlandini$^2$ and A L Stella$^3$}
\address{$^1$ Dipartimento di Fisica and  sezione CNISM,
Universit\`{a} di Padova, Padova, Italy}\ead{baiesi@pd.infn.it}
\address{$^2$ Dipartimento di Fisica and  sezione CNISM,
Universit\`{a} di Padova, Padova, Italy}\ead{orlandini@pd.infn.it}
\address{$^3$ Dipartimento di Fisica, and sezione INFN,
 Universit\`{a} di Padova, Padova, Italy}\ead{stella@pd.infn.it}

\begin{abstract}
We estimate by Monte Carlo simulations the configurational entropy of $N$-steps polygons in the
cubic lattice with fixed knot type. By collecting a rich  statistics of configurations with very
large values of $N$ we are able to analyse the asymptotic behaviour of the partition function of
the problem for different knot types. Our results confirm that, in the large $N$ limit, each prime
knot is localized in a small region of the polygon, regardless of the possible presence of other
knots. Each prime knot component may slide along the unknotted region contributing  to the overall
configurational entropy with a term proportional to  $\ln N$. Furthermore, we discover that the mere
existence of a knot requires a well defined entropic cost that scales exponentially with its
minimal length.  In the case of polygons with composite knots it turns out that the partition
function can be simply factorized in terms that depend only on prime components  with an additional
combinatorial factor that takes into account the statistical property that by interchanging two identical
prime knot components in the polygon the corresponding set of overall configuration remains unaltered. 
Finally, the above results allow  to  conjecture a sequence of inequalities for the connective constants of
polygons whose topology varies within a given family of composite knot types.
\end{abstract}

\submitto{JSTAT}

\maketitle

\section{Introduction}
\setcounter{equation}{0}

A long, flexible polymer chain in good solvent can be highly self-entangled~\cite{DeGennes:1979,Doi&Edwards:1986}
and, if a ring closure reaction occurs, or if its extremities are hold tied by some device, the
entanglement can be trapped as a knot~\cite{Vologodskii:1974:JETP,Orlandini&Whittington:2007:Rev-Mod_Phys}.
Moreover, because of the excluded volume interaction, a
knotted molecule cannot change its topological status, without breaking and reconnecting back
chemical bonds. This for example is the situation  one encounters in biological systems where
special enzymes, called topoisomerases, can pass one strand of the double stranded circular DNA through
another and knot or unknot the molecule, to facilitate elementary cellular
processes~\cite{Sumners:1995,Wang:1996:AnnRevBio}.
In general, however, there is no spontaneous transition between different knotted statuses and
in the most common experimental situations the topology of the ring does not change in time. Clearly, the
presence of topological constraints limits the configurational space available to the ring, with a
consequent reduction of the entropy of the system, compared to the topologically unconstrained
case~\cite{Sumners&Whittington:1988:J-Phys-A}. It is then interesting to precisely quantify this entropy loss and
to determine how it depends on the particular topology (i.e.~knot type) considered.

Unfortunately, most of the theoretical studies performed so far refer to  the ensemble in which the
rings may assume all the topologies. The reason is that polymer rings in good solvent can be
modelled as self-avoiding polygons (SAPs or simply polygons), which are in turn mapped to a magnetic system at its
critical point and studied by renormalization group techniques~\cite{DeGennes:1979,Guida&Zinn-Justin:1997:JPA,Vanderzande:1998}. This approach has led to the well
established result that the number of  $Z(N)$ of $N$-steps SAPs grows, for large $N$, as
\begin{equation}
Z(N)\simeq A \mu^N N^{\alpha-2} \label{Zfree}
\end{equation}
where the amplitude $A$ and the connective constant $\mu$ are non-universal quantities that depend on the microscopic
features of the chain while $\alpha $ is a universal exponent given by $\alpha = 2-d \nu$, where
$d$ is the dimensionality of the space and $\nu$ the metric exponent \cite{Vanderzande:1998}.
In $d=3$ dimensions, numerical simulations~\cite{Clisby:2010:PRL} give for self
avoiding loops the estimate $\nu \simeq 0.587597(7)$,
and consequently $\alpha \simeq 0.237209(21)$, in agreement with
field theoretical results~\cite{Guida&Zinn-Justin:1997:JPA}.
Since for the subset of SAPs with a given knot type $k$ the above mentioned mapping is not valid anymore,
there is no field theory argument to establish a scaling similar to ($\ref{Zfree}$) for $Z_k(N)$.
However, it is reasonable to expect that
\begin{equation}
Z_{k}(N) \simeq A_{k} \mu_{k}^N N^{\alpha_{k}-2}
\label{Zk}
\end{equation}
where $\mu_{k}$ and $\alpha_k$ are, respectively, the connective constant and the entropic exponent
of the subset of SAPs with fixed knot type $k$. For a generic knot type $k$ there is no rigorous
relation between $\mu_k$ and $\mu$ but in the case of \emph{unknotted polygons} (i.e.~SAPs with
trivial topology, $k=\emptyset$) it is possible to prove rigorously that $\mu_{\emptyset} < \mu$
\cite{Sumners&Whittington:1988:J-Phys-A} whereas numerical estimates of $\alpha_{\emptyset}$
suggests the intriguing identity $\alpha \simeq \alpha_{\emptyset}$ \cite{Orlandini:1998:J-Phys-A},
although results presented so far are not sharp enough to rule out completely a possible, although
small, discrepancy between the two entropic exponents, i.e. $\alpha_{\UN}\simeq\alpha$. 
One among the results presented here concerns the improvement of the estimate 
$\alpha-\alpha_{\emptyset}$ and of the ratio $A_\UN/A$ (see section~\ref{sec:MC}),this one performed, to
our knowledge, for the first time.

Note that Eq.~(\ref{Zfree}) and (\ref{Zk}) with $k=\emptyset$ have interesting implications for the
probability of realizing an unknot in the ensemble with
unrestricted topology, $P_{\emptyset}(N) \equiv Z_{\emptyset}(N)/ Z(N)$. Indeed, from (\ref{Zfree}) and
(\ref{Zk}) with $k=\emptyset$ one gets
\begin{equation}
P_\UN\sim \frac{A_\UN}{A} \left(\frac {\mu_\UN}{\mu}\right)^N N^{\alpha_\UN - \alpha} = \frac{A_\UN}{A} e^{-N/N_0}
N^{\alpha_\UN - \alpha} \label{un_prob}
\end{equation}
and since $\mu_{\emptyset} < \mu$ we get the well known result that the unknotting probability goes
to zero exponentially fast with $N$ \cite{Sumners&Whittington:1988:J-Phys-A}.
The parameter $N_0=1/\log(\mu/\mu_\UN)$ gives a typical
number of steps above which the unknot probability is reasonably low or, in other words,
the occurrence of knots is not negligible anymore. Previous numerical estimates for polygons on the cubic
lattice gave $N_0 \approx 2\times 10^5$
\cite{Janse-van-Rensburg&Whittington:1990:J-Phys-A,Janse-van-Rensburg:2002, Rensburg&Rechnitzer:2008:JPA}.

Since for polymer rings with a generic, fixed knot type $k$ neither analytical tools nor rigorous arguments are
available, one has to rely entirely on numerical approaches and scaling arguments in the analysis of the
above issues.  By using
the BFACF algorithm~\cite{Berg81,Aragao83} (the acronym comes from the initials of the authors)
coupled to a multiple Markov chain sampling technique, and assuming for SAPs with fixed knot
type $k$ the scaling (\ref{Zk}), evidence found \cite{Orlandini:1996:J-Phys-A} that
\begin{equation}
\mu_k=\mu_{\UN} \qquad \alpha_k = \alpha_{\UN} + \pi_k, \label{conj}
\end{equation}
where $\pi_k$ is the number of prime components in the knot decomposition of $k$.
It is interesting to notice that results similar to (\ref{conj}) have been obtained also for off-lattice
models of rings such as the bead-rod models \cite{Deguchi&Tsurusaki:1997:PRE}
suggesting that the scaling  behaviour (\ref{Zk}) with (\ref{conj})
is a universal property of loops in free space with a given  knot type $k$.
Relations in (\ref{conj}) are consistent with recent findings showing that prime knots in swollen rings are
weakly localized, i.e.~have an average ``length'' $\langle l \rangle \sim N^{t}$ with an exponent $0<t<1$, which
has been estimated in
\cite{Marcone:2005:J-Phys-A,Marcone:2007:PRE,Orlandini:2009:Phys-Bio} as $t \simeq 0.7$.
Indeed weak localization of prime knots implies that, in
the limit $N\to\infty$, each prime component behaves essentially as a decorating vertex fluctuating
along the unknotted ring. This additional configurational degree of freedom brings a factor $N$ in
front of $Z_{\UN}$ for each prime component and, in the general case of  a knot $k$ made by $\pi_k$ prime
components, one may guess:
\begin{equation}
Z_k(N) \simeq  N^{\pi_k} Z_{\emptyset}(N). \label{Zkun}
\end{equation}
Although the above simple argument furnishes a plausible explanation of  relations (\ref{conj}), it
is too crude to fully characterize the entropy of a knotted ring even in the large $N$ limit.
For example, the amplitude $A_k$ is still undetermined and there is no trace of the type of
prime knots that contribute to $Z_k(N)$. In fact, by regarding prime knots as point-like
objects, we are neglecting the effective entropic cost that the system has to pay in order
to tie them into unknotted loops.
This entropic cost would decrease the $Z_k(N)$ in
(\ref{Zkun}) by a factor, say, $C_k$ giving the more precise  expression
\begin{equation}
Z_{k}(N)\simeq \frac{A_\UN}{C_k} \mu_\UN^N N^{\alpha_\UN-2+\pi_k}. \label{mor_det_Zk}
\end{equation}
It is interesting to notice that, if $C_k$ is related to a sort of entropic cost to pay in order to tie
a given knot, its value should depend on the  knot type $k$ and not only on the number of its prime components.
If this is the case,  Eq. (\ref{mor_det_Zk}) would furnish a more fundamental description of
the large $N$ behavior of the entropy of  a knot since it would distinguish the type of knot hosted
by the polygon.
This description should depend also on topological invariants of $k$ other than  $\pi_k$.

It is important to stress that a numerical check of the validity of (\ref{mor_det_Zk}) and,
in particular, a numerical estimate of
$C_k$ as a function of $k$, is a quite hard task to perform because it requires a good statistics of
polygon configurations with very large values of $N$. This is particularly crucial for SAPs
on discrete lattices, for which a reasonable amount of knotted configurations can be sampled only
for $N \ge N_0 \sim 10^5$.
This is probably the reason why no attempts have been made so far to look in more details
at the asymptotic form (\ref{mor_det_Zk}). In this paper we explore this issue by sampling
polygons on the cubic lattice with $N$ up to $200000$. Unlike in
previous Monte Carlo's, where the sampling was performed in the fixed knot ensemble using BFACF
~\cite{Berg81,Aragao83} algorithm, we decided to sample in the free topology ensemble by using
the very efficient two-pivot-points algorithm~\cite{Madras&Slade} and subsequently to partition the sampled
configurations according to their topology

In section~\ref{sec:MC} we describe the algorithm that we use to sample
knotted SAPs and the procedure designed to detect knots out of configurations that,
for large values of $N$, turn out to be highly intricated.
As a first outcome of this investigation we will give a sharper estimate both of the difference
$\alpha-\alpha_0$ and of the ratio $A_k/A$. This will establish a more detailed relation between
the subclass of unknotted rings and the full class of rings with
unrestricted topology. In section~\ref{sec:Ck} we test the validity of
(\ref{mor_det_Zk}) and estimate $C_k$ as a function of $k$. This is the main result of the paper:
it will be first established for the simplest case of prime knots and later generalized to
composite knots.
Section~\ref{sec:Ck:mu} also includes further conjectures on the connective constants of
SAP ensembles with restricted topology.
Section~\ref{sec:end} is devoted to discussion and conclusions.

%%%%%%%%%%%%%%%%%%%%%%%%%%%%%%%%%%%%%%%%%%%%%%%%%%%%%%%%%%%%%%%%%%%
\begin{figure}[!tb]
\begin{center}
\includegraphics[angle=0,width=11cm]{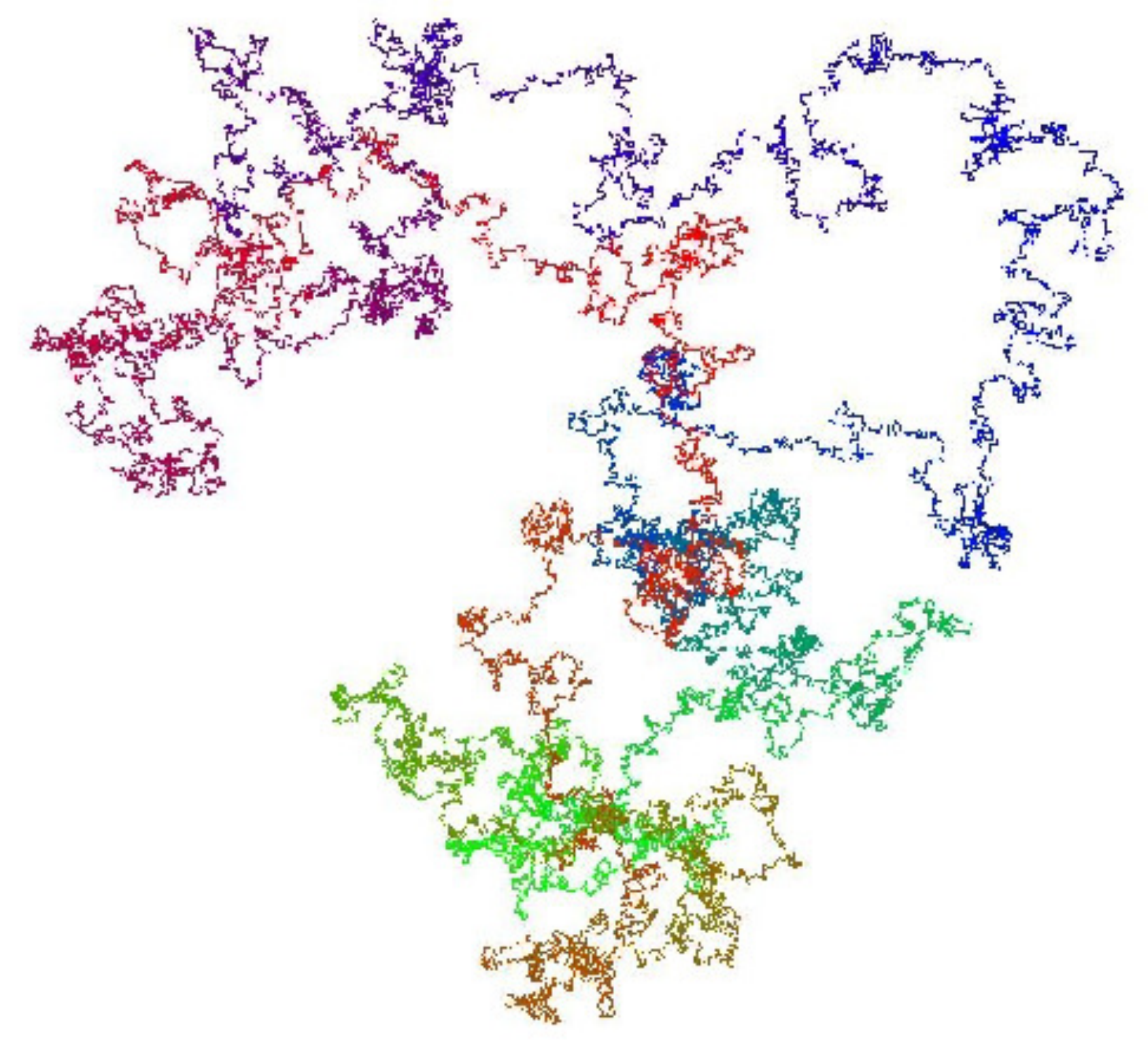}
\vskip 3mm
\includegraphics[angle=0,width=10.5cm]{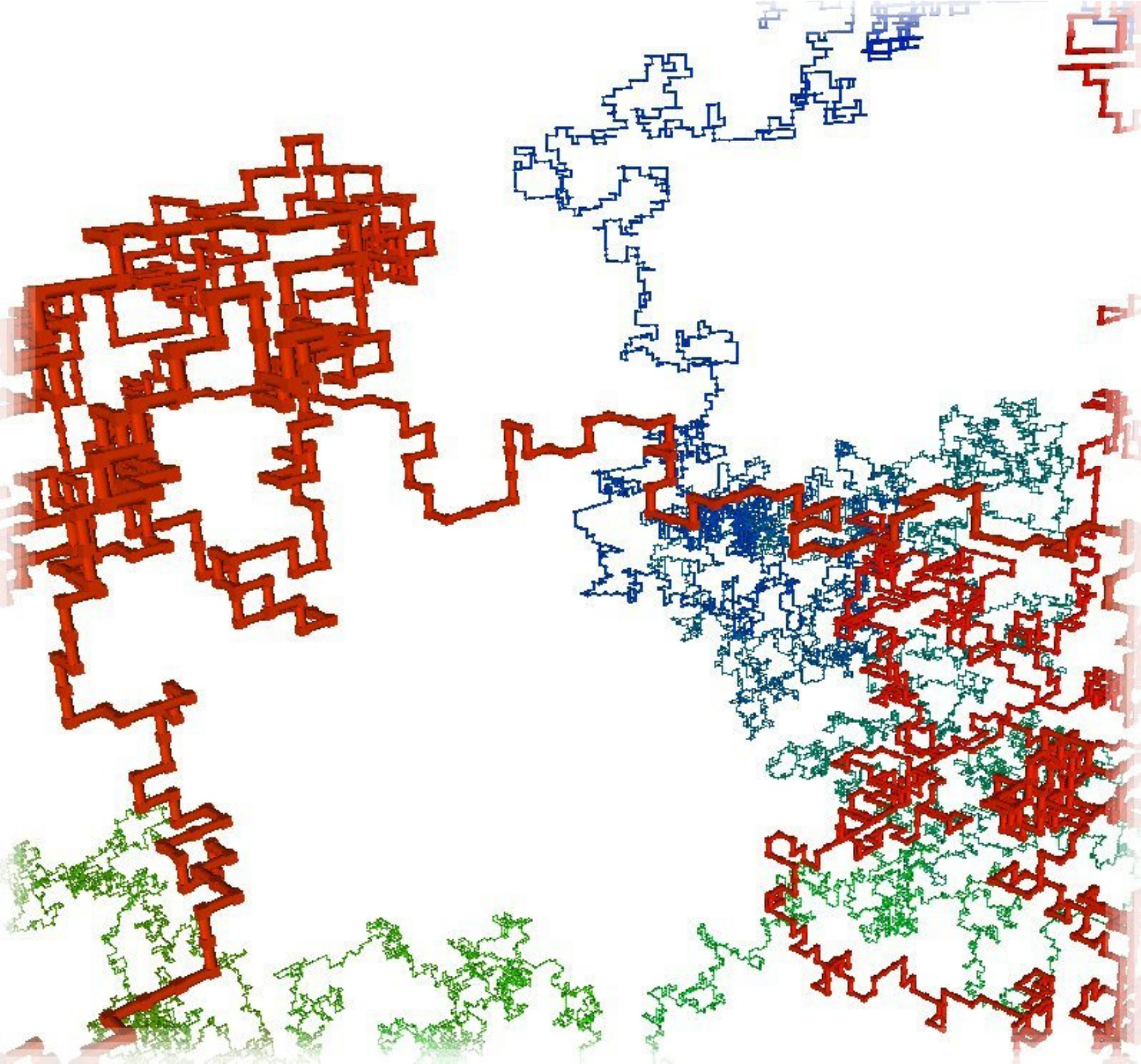}
\vskip 3mm \caption{Equilibrium configuration of SAP on the cubic lattice with $N=50000$ steps (above)
and a detail of its central part (below).\label{fig:conf}}
\end{center}
\end{figure}
%%%%%%%%%%%%%%%%%%%%%%%%%%%%%%%%%%%%%%%%%%%%%%%%%%%%%%%%%%%%%%%%%%%

\section{Model, Monte Carlo method and knot detection procedure}\label{sec:MC}
To model polymer rings with excluded volume interaction we consider $N$-step SAPs on the cubic lattice,
i.e.~self-avoiding walks with the two extremities that are one lattice distant apart.
These polygons are sampled in free space by using the two-pivot moves,
a fixed-$N$ algorithm that has been proved to
be ergodic in the class of all polygons and shown to be very efficient in sampling uncorrelated
configurations~\cite{Madras&Slade}. With this procedure we generate configurations with $N$ up to
$200000$.
As an example, in Fig.~\ref{fig:conf} we plot a configuration with $N=50000$, together with a
closer view of part of it.

%%%%%%%%%%%%%%%%%%%%%%%%%%%%%%%%%%%%%%%%%%%%%%%%%%%%%%%%%%%%%%%%%%%
\begin{figure}[!tb]
\begin{center}
\includegraphics[angle=0,width=7cm]{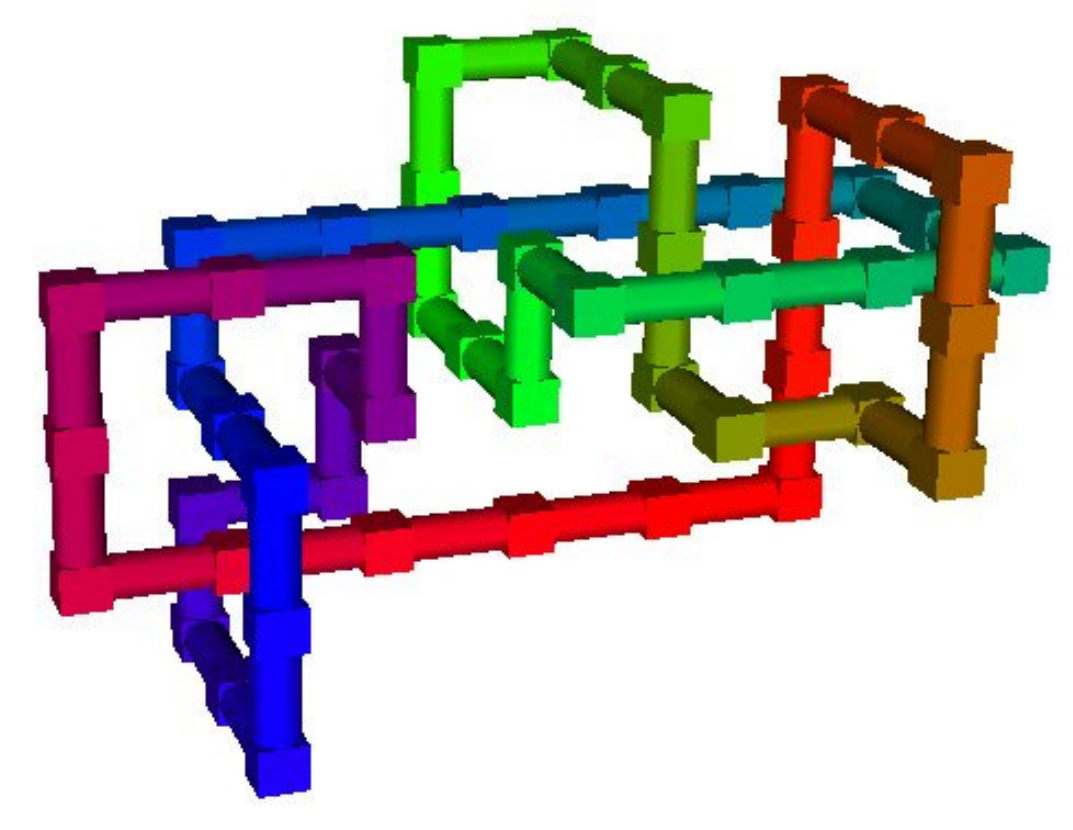}
\includegraphics[angle=0,width=6cm]{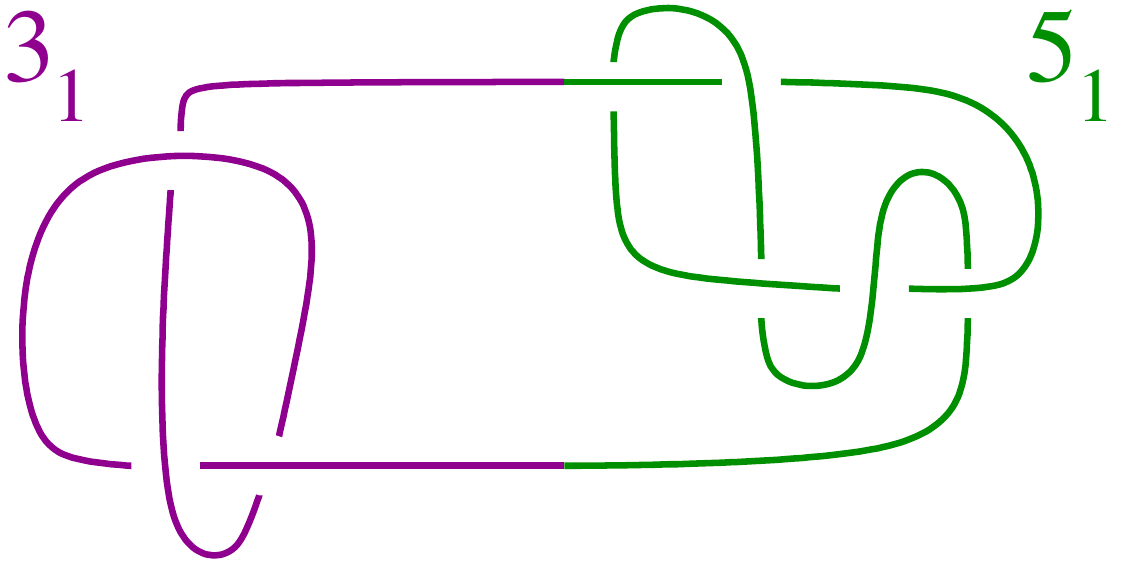}
\vskip 3mm \caption{By applying the step-reduction algorithm, based on the BFACF algorithm, that
preserves the topology the configuration of Fig.~\ref{fig:conf} is simplified to the one
represented in top panel and having $N=82$ steps. A further simplification of the Dowker code
allows to identify the knot as the composite knot $3_1\#5_1$ knot whose minimal diagram
representation is shown on the right.
\label{fig:shrink}}
\end{center}
\end{figure}
%%%%%%%%%%%%%%%%%%%%%%%%%%%%%%%%%%%%%%%%%%%%%%%%%%%%%%%%%%%%%%%%%%%

Since the pivot moves can change the knot type of polygons, the topology of each configuration must
be detected by means of some topological invariant. This is indeed the most problematic part of the
whole investigation since, even in good solvent conditions,  very long polygons may assume a very
intricated spatial arrangement. This ``geometrical'' entanglement gives rise, in general, to knot
projections with a very large number of unessential crossings (from the topological point of view)
that severely hinders the knot detection algorithm based on the calculation of polynomial
invariants ~\cite{Orlandini:2007:Rev-Mod_Phys}.

To circumvent this difficulty, we simplify each sampled configuration before performing its
planar projection. This is achieved by applying to the polygon a smoothing algorithm that
progressively reduces the length of the chain while keeping its knot type unaltered (for a similar
procedure, see ~\cite{Micheletti:2006:J-Chem-Phys:16483240,Baiesi:2007:Phys-Rev-Lett:17930800,
Baiesi_et_al:2009:JCP}). This procedure is based on the
$N$-varying BFACF algorithm ~\cite{Berg81,Aragao83} and has the nice feature of being ergodic
within each knot type. We set a sufficiently small step fugacity (i.e. the parameter conjugate to
$N$), such that the algorithm induces a rapid reduction in the number of steps of the polygon. This
simplification technique can reduce dramatically the number of crossings encountered in an
arbitrary projection. An example of how efficient this simplification procedure  can be is shown in
Fig.~\ref{fig:conf} where a configuration of initially $N=50000$ steps is shrunk  down to the
$N=82$ steps configuration of Fig.~\ref{fig:shrink}. A further reduction is achieved by performing
$500$ projections and choosing the projection with the minimal number of crossings. The resulting
knot diagram is encoded in terms of the Dowker code~\cite{Adams:1994}. A further simplification of
the Dowker code based on Reidemeister-like moves is performed. Finally, a factorization of the
Dowker code is attempted. This procedure, whenever successful, splits composite knots into their
prime components. From each component of the original Dowker code we extract, by using {\sc
Knotfind}~\cite{knotfind}, the knot type of the original configuration (see Fig.~\ref{fig:shrink}
for the example given in Fig.~\ref{fig:conf}). In this way we have been able to distinguish
composite knots with up to $5$ prime components, and with each component having crossing number up
to 11 \cite{Baiesi:2007:Phys-Rev-Lett:17930800,Baiesi_et_al:2009:JCP}. The unbiased sampling with
unconstrained topology allows us to estimate the probability $P_{k} \equiv Z_{k} / Z$ of occurrence
of  a given knot type $k$ and to estimate its configurational entropy with respect to unknotted
polygons, i.e.~the ratio $Z_{k} / Z_{\UN}$. Since the statistics of unknotted polygons will be used
extensively as a reference, it is convenient to start by performing a good estimate of
$Z_\UN(N)$. This can be achieved by looking at the scaling of the unknotting probability
(\ref{un_prob}) as $N$ increases.

%%%%%%%%%%%%%%%%%%%%%%%%%%%%%%%%%%%%%%%%%%%%%%%%%%%%%%%%%%%%%%%%%%%
\begin{figure}[!tb]
\begin{center}
\includegraphics[angle=0,width=11cm]{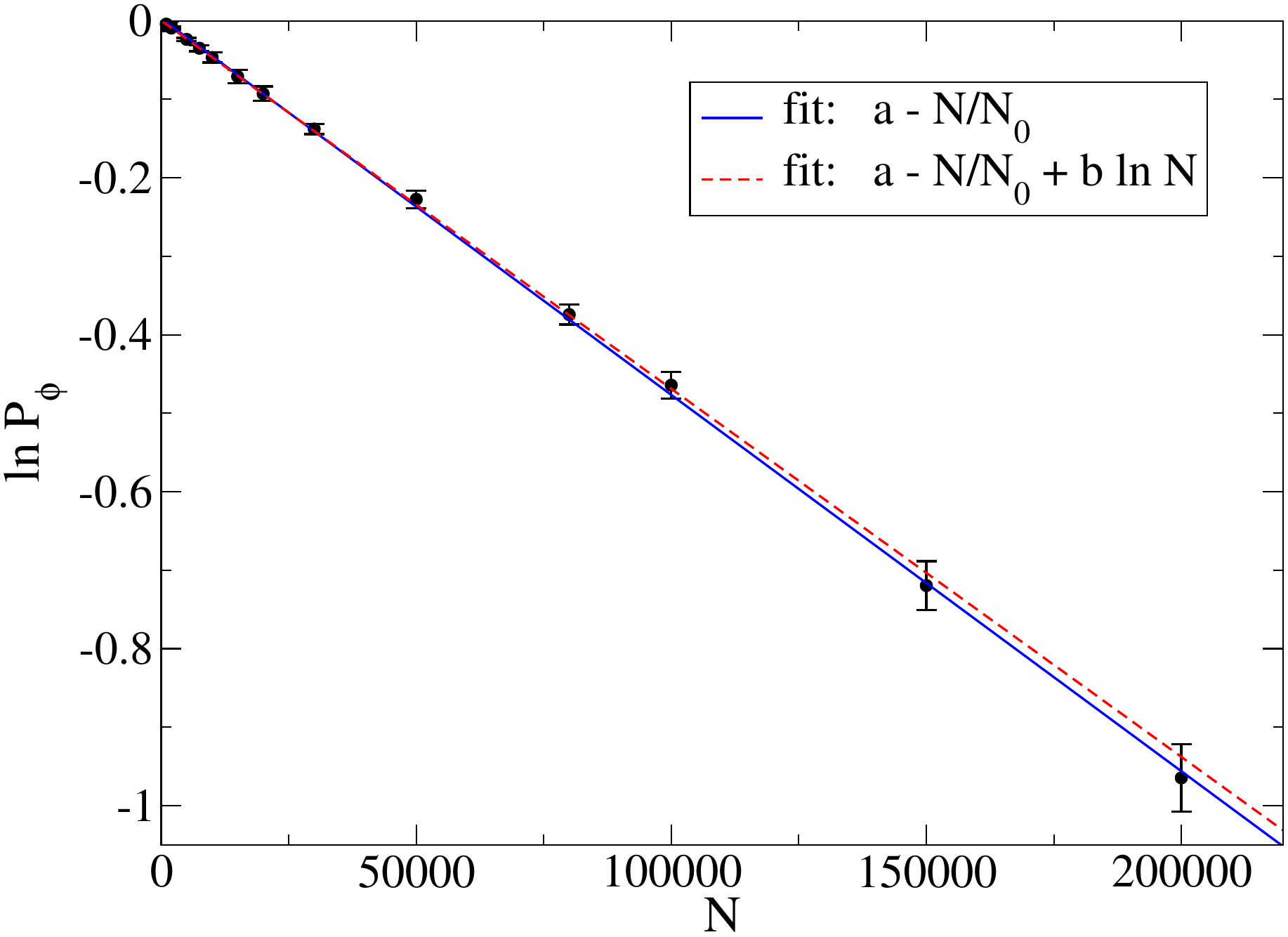}
\vskip 3mm
\caption{Decay of the probability of unknotted configurations (log-scale) with the chain length.
Fits are also shown.
\label{fig:No}}
\end{center}
\end{figure}
%%%%%%%%%%%%%%%%%%%%%%%%%%%%%%%%%%%%%%%%%%%%%%%%%%%%%%%%%%%%%%%%%%%

In Fig.~\ref{fig:No} we plot $\ln P_\UN$ as a function of $N$. The two lines correspond to two
different fits of the data. To estimate the difference $\alpha-\alpha_0$ we first perform a nonlinear
fitting (dashed line) of the form  $a -N/N_0 + b \ln N$. This  yields $\alpha-\alpha_0 =
b=-3\times10^{-5}\approx 0$ confirming the conjecture $\alpha=\alpha_0$. If we now
assume $\alpha_\UN=\alpha$ we can perform a linear fit (solid line) $a-N/N_0$. This gives  $N_0 =
210400 \pm 1300$ and $a=0.003(2)$. The estimate of $a$ strongly suggests that within error bars
$A_\UN=A$. This last result is  quite interesting since it strengthens the relation between
the statistics of unknotted SAPs and the one of all SAPs, not only at the level of the
entropic exponents, but also at the level of amplitudes.
Clearly the main difference  relies on the entropies per monomer $\mu$ and $\mu_\UN$.
However, the difference $\mu-\mu_\UN\simeq \mu / N_0$ is very small:
with the most recent and precise estimate $\mu=4.684044 \pm 0.000011$ by
Slade and coworkers~\cite{Clisby:2007:JPA},
we estimate $\mu-\mu_\UN=0.000022(2)$ (that is twice the statistical error for $\mu$)
and thus $\mu_\UN = 4.684022\pm  0.000013$.

\section{The entropic cost of a knot}\label{sec:Ck}

\subsection{Prime knots}

%%%%%%%%%%%%%%%%%%%%%%%%%%%%%%%%%%%%%%%%%%%%%%%%%%%%%%%%%%%%%%%%%%%
\begin{table}[!b]
\begin{center}
\begin{tabular}{|c|c|c|c|c|c|}
\hline
\hfil knot \hfil& $\alpha_k-\alpha$  \hfil & \hfil $C_k$ \hfil &  \hfil $V_k=\log_{\mu_\UN} C_k$ \hfil & \hfil $\ell_k$ \hfil \\
\hline
$3_1$& 1.002(7)& $227800\pm 1400$ &7.989(4)&24\\
$4_1$& 0.96(3)& $5.04(15)\times 10^6$  &9.995(20)&30\\
$5_1$& 1.13(6)& $4.48(35)\times 10^7$  &11.41(5)&34\\
$5_2$& 1.10(8)& $3.19(25)\times 10^7$  &11.19(5)&36\\
$6_1$& 1.23(25)& $60(24)\times 10^7$ &13.1(2) &40\\
$6_2$& 1.22(13)& $38(12)\times 10^7$ &12.8(2) &40\\
$6_3$& 1.08(22)& $61(19)\times 10^7$ &13.1(2) &40\\
\hline
\end{tabular}
\caption{Estimates of the difference $\alpha_k-\alpha_\UN$ (second column) and of the entropic cost $C_k$ (third column) for the simplest prime knots. The third column suggest a simple relation
between the entropic cost $C_k$ and the the minimal length $\ell_k$ necessary to tight the
prime knots of the fist column as estimated in~\cite{vanR_minlength:1995:JKTR}.
\label{tab:1}}
\end{center}
\end{table}
%%%%%%%%%%%%%%%%%%%%%%%%%%%%%%%%%%%%%%%%%%%%%%%%%%%%%%%%%%%%%%%%%%%

\begin{figure}[!tb]
\begin{center}
\includegraphics[angle=0,width=11cm]{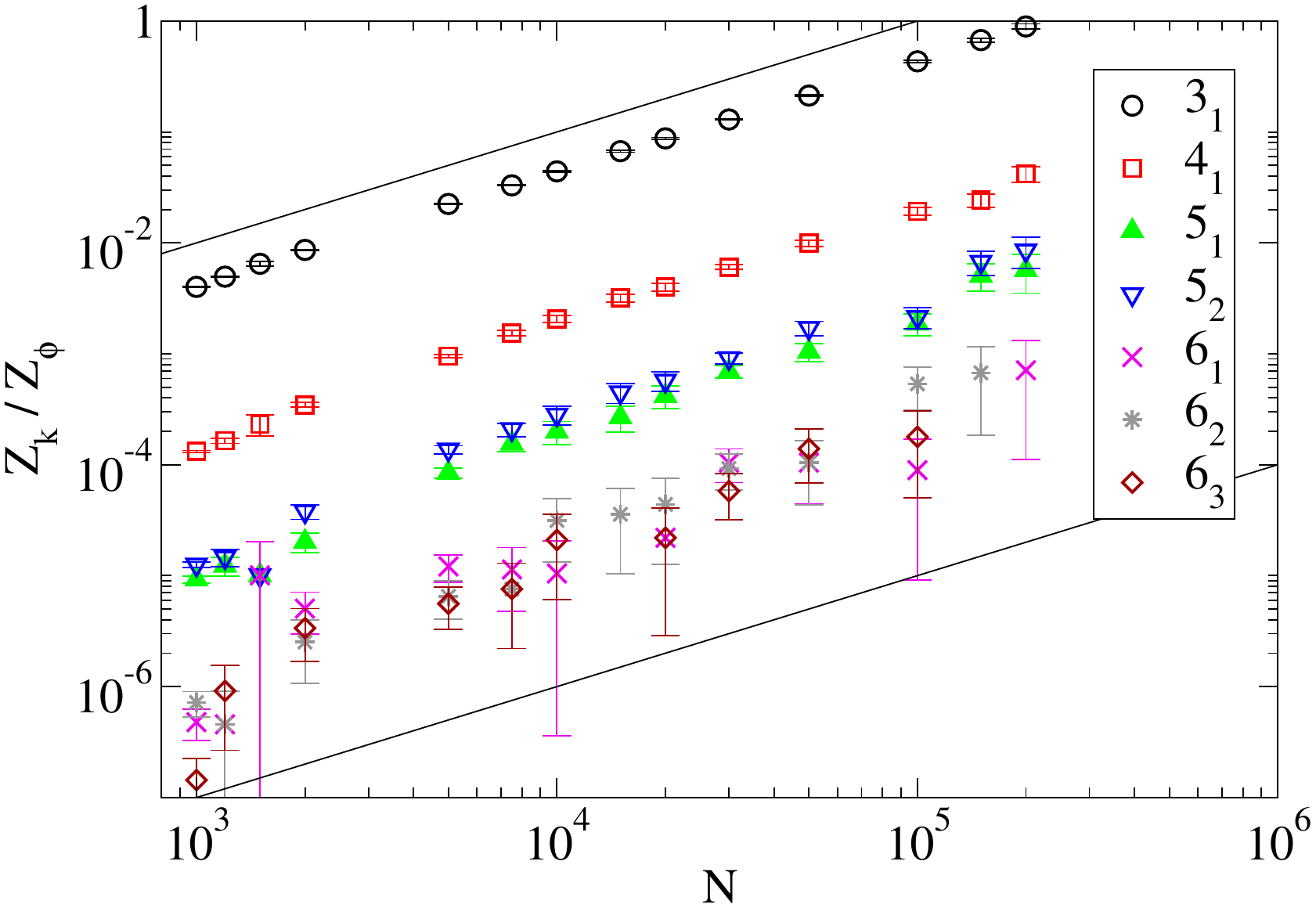}
\vskip 3mm
\caption{Partition function of the simplest prime knots
divided by that of the unknot (log-log scale vs $N$).
Straight lines are a guide to the eye, scaling $\sim N$.
\label{fig:345}}
\end{center}
\end{figure}

To estimate the entropic cost $C_k$ for polygons with fixed knot type $k$, we compute the ratios
$Z_k(N)/Z_\UN(N)$. Indeed, by assuming the scaling form (\ref{mor_det_Zk})  we expect
$Z_k(N)/Z_\UN(N)\simeq N^{\pi_k}/ C_k$. Let us consider first knotted SAPs where $k$ is a prime knot.
Fig.~\ref{fig:345} shows the $N$-behavior (in log-log scale) of the ratio
$Z_k(N)/Z_\UN(N)$ for prime knots up to $6$-crossings. As expected from (\ref{mor_det_Zk}), no
exponential behavior is observed and the scaling $ \sim N / C_k$ is confirmed (note that
$\pi_k=1$ since we are considering prime knots), with a $C_k$ whose value increases as the knot
complexity increases. The estimates of $\alpha_k-\alpha_\UN$, reported in Table~\ref{tab:1} in the
second column, are all consistent with the relation $\alpha_k-\alpha_\UN=1$. The estimates of $C_k$
are reported in the third column of Table~\ref{tab:1}. The most striking feature to notice is the
simple relation observed between the value of $C_k$ and the knot type $k$. Indeed,  from column
$4$, it turns out that, to a good approximation, the entropy cost necessary to host a prime knot
$k$ goes like
\begin{equation}
C_k \simeq \mu_\UN^{\ell_k /3}
\end{equation}
where $\ell_k$ (see last column of Table~\ref{tab:1}) is the minimal length required to tie a knot $k$
on the cubic lattice~\cite{vanR_minlength:1995:JKTR}.
Thus, the entropic cost $C_k$ is intimately related to a ``microscopic'' property of the knot $k$,
that is, the length of its ``ideal''
representation~\cite{Katritch_et_al:1996:Nature,Katritch:1997:Nature:9217153} on the cubic
lattice~\cite{vanR_minlength:1995:JKTR}.

%%%%%%%%%%%%%%%%%%%%%%%%%%%%%%%%%%%%%%%%%%%%%%%%%%%%%%%%%%%%%%%%%%%
%%%%%%%%%%%%%%%%%%%%%%%%%%%%%%%%%%%%%%%%%%%%%%%%%%%%%%%%%%%%%%%%%%%

It is tempting to interpret $V_k= \ell_k/3$ as an equivalent number of monomers ``lost'' by the
polygon in order to form the knot. For example, the partition function of a trefoil, $Z_{3_1}$, would be
described by the exponential factor $\mu_\UN^{N-V_{3_1}}$ with $V_{3_1} = 8$. In other words, a
$N$-step polygon with a $3_1$ knot has the same configurational entropy (in the limit of large $N$) as
an unknotted polygon with $N - 8$ steps endowed with a sliding decorating vertex (the knot).
These findings suggest that it is sufficient to  know the length $\ell_k$
of a given prime knot in its ideal lattice representation in order to make an accurate prediction of its
frequency along a swollen  ring.

The factor of $1/3$ is quite intriguing and we have no explanation for that so far. Clearly it will
be important to test further this value by looking at more complicated knots. This would require a
much larger statistics and consequently much larger values of $N$. Another interesting issue
would be to see if the relation is model-dependent by looking for example at polygons embedded on
different lattices.

\subsection{Composite knots}
We now extend the analysis of $C_k$ to composite knots, namely knots made by connecting
 prime component knots (such as the $3_1\#5_1$ in Fig.~\ref{fig:shrink}).
In the most general case we may assume $k$ to be composed by the prime knots
$k_1,k_2,\cdots,k_m$, each appearing respectively $\pi_{1}$, $\pi_{2}$ and  $\pi_{m}$ times. The
number $\pi_{i}$ represents somehow the degree of degeneracy of the knot prime $k_i$ in the
composite knot $k$. For the composite knot $k$ the entropic cost $C_k$ could, in principle, depend
on the set $\{k_i\}$ in a quite complicated way. However, if we still assume that, in the large $N$
limit,  each prime knot localizes along the chain, regardless of the presence of other knot components,
we can make the working hypothesis that the cost of a composite knot $k_1\#k_2$ factorizes as
$C_{k_1\#k_2} = C_{k_1}\times C_{k_2}$ (for $k_1\ne k_2$). This will bring to the conjecture that,
in the $N\to\infty$ limit,
\begin{equation}
Z_{k}(N) \simeq Z_{\UN}(N) \left[
\frac{1}{(\pi_{1})!}\left(\frac{N}{C_{k_1}}\right)^{\pi_{1}} \!\cdots
\frac{1}{(\pi_{m})!}\left(\frac{N}{C_{k_m}}\right)^{\pi_{m}}
\right].
\label{eq:0}
\end{equation}
The presence of the factorial terms $1/(\pi_{i})!$ can be explained as follows: if in a polygon
with a given knot type there are a number $\pi_{i}$  of the same prime knot $k_i$, then if we
permute the position of these prime knots in the polygon the set of corresponding overall 
configurations remains the
same. Since in the original statistics we do not take care of this over-counting  we have to divide
the original partition function by $1/(\pi_{i})!$, and this must be done for any prime knot
component in the knot decomposition.

With these notations, the full cost of a composite knot is then
\begin{equation}
C_k = \prod_i \pi_i!\; C_{k_i}^{\;\pi_{i}}
\label{eq:Ck}
\end{equation}
Eqs.~(\ref{eq:0})-(\ref{eq:Ck}) suggests that, if we knew the entropic cost $C_{k_i}$ necessary to
tie  of each prime component $k_i$,  the number of configurations $Z_{k}(N)$ of the composite knot
$k$ could be easily deduced, in the large $N$ limit, by looking at the partition function
$Z_{\UN}(N)$ of unknotted polygons of the same length.

%%%%%%%%%%%%%%%%%%%%%%%%%%%%%%%%%%%%%%%%%%%%%%%%%%%%%%%%%%%%%%%%%%%
\begin{figure}[!tb]
\begin{center}
\includegraphics[angle=0,width=11cm]{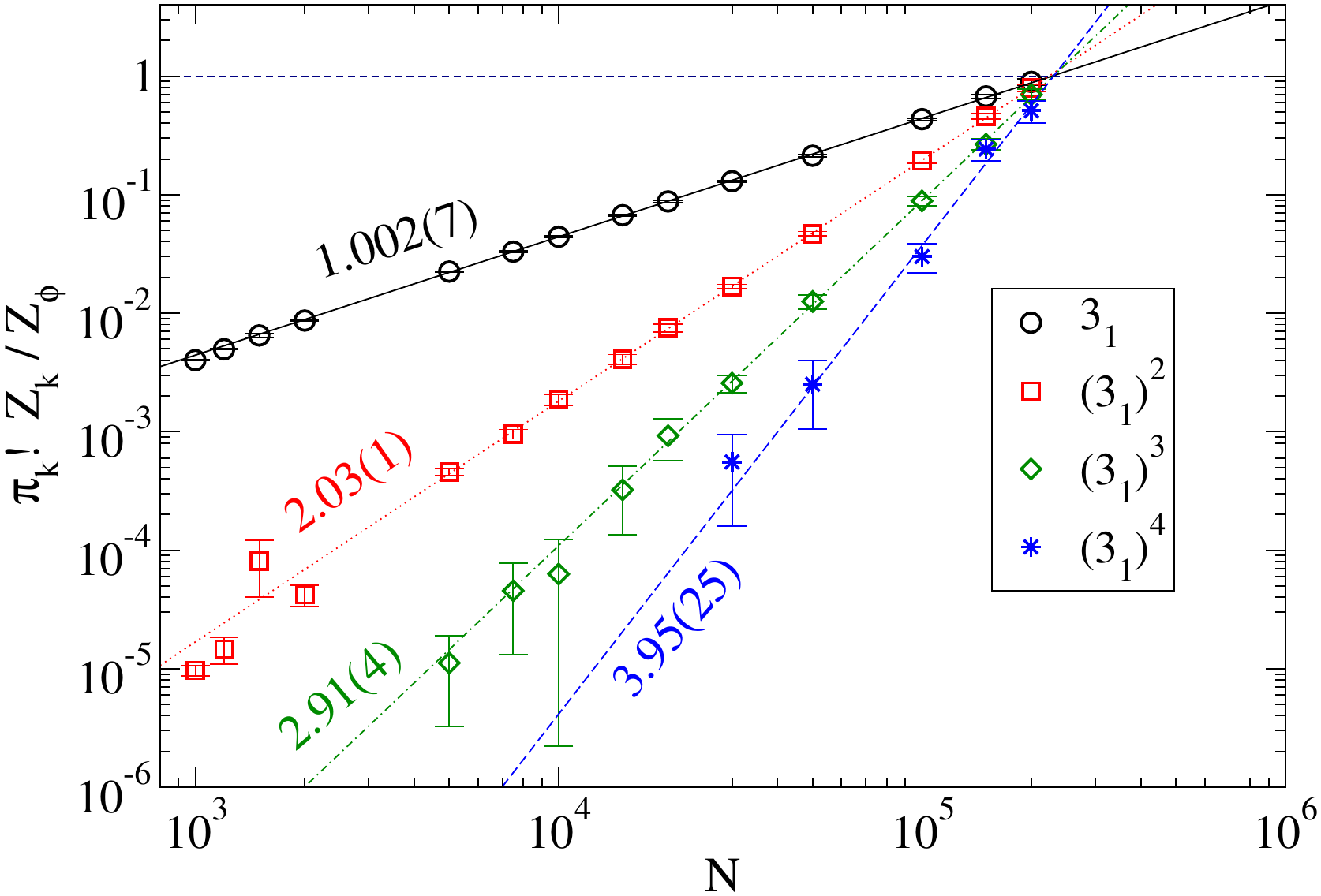}
\vskip 3mm
\caption{Partition function of knots involving copies of $3_1$ divided
by that of the unknot and multiplied by the factorial of the number
of $3_1$ components, in log-log scale as a function of the chain length.
Straight lines are power-law fits (exponents are shown close to them).
\label{fig:many31}}
\end{center}
\end{figure}
%%%%%%%%%%%%%%%%%%%%%%%%%%%%%%%%%%%%%%%%%%%%%%%%%%%%%%%%%%%%%%%%%%%

We first check Eq.~(\ref{eq:0}) for composite knots including only copies of the trefoil knot,
for which we have good statistics up to four prime components.
From  Eq.~(\ref{eq:0})  we expect the following relation to hold:
\begin{eqnarray}
Z_{3_1} / Z_{\UN}           & \simeq & N / C_{3_1}
\nonumber\\
Z_{3_1\#3_1} / Z_{\UN} \times 2     & \simeq & (N / C_{3_1})^2
\label{eq:31} \\
Z_{3_1\#3_1\#3_1} / Z_{\UN} \times 3! & \simeq & (N / C_{3_1})^3
\nonumber\\
Z_{3_1\#3_1\#3_1\#3_1} / Z_{\UN} \times 4! & \simeq & (N / C_{3_1})^4
\nonumber
\end{eqnarray}
In Fig.~\ref{fig:many31} we show these ratios times the suitable factorials, in log-log scale as a
function of $N$. The four straight lines are power-law fits whose exponents agree within error bars
with Eqs.~(\ref{eq:31}). Moreover, as expected, all fits cross each other at a single point $(C_{3_1},
1)$ with  abscissa $C_{3_1}\approx 227000$. Hence the starting assumption that the total entropic
cost to tie a composite knot of $\pi_i$ prime knots simply factorizes (see Eq. \ref{eq:Ck}) is
crisply verified, at least for trefoil knots. Note that, by extrapolating the data of
Fig~\ref{fig:many31} to larger values of $N$ it is clear that for $N > C_{3_1}$ it is more
convenient, entropically, to tie composite knots made by trefoils than forming unknotted polygons.

%%%%%%%%%%%%%%%%%%%%%%%%%%%%%%%%%%%%%%%%%%%%%%%%%%%%%%%%%%%%%%%%%%%
\begin{figure}[!tb]
\begin{center}
\includegraphics[angle=0,width=11cm]{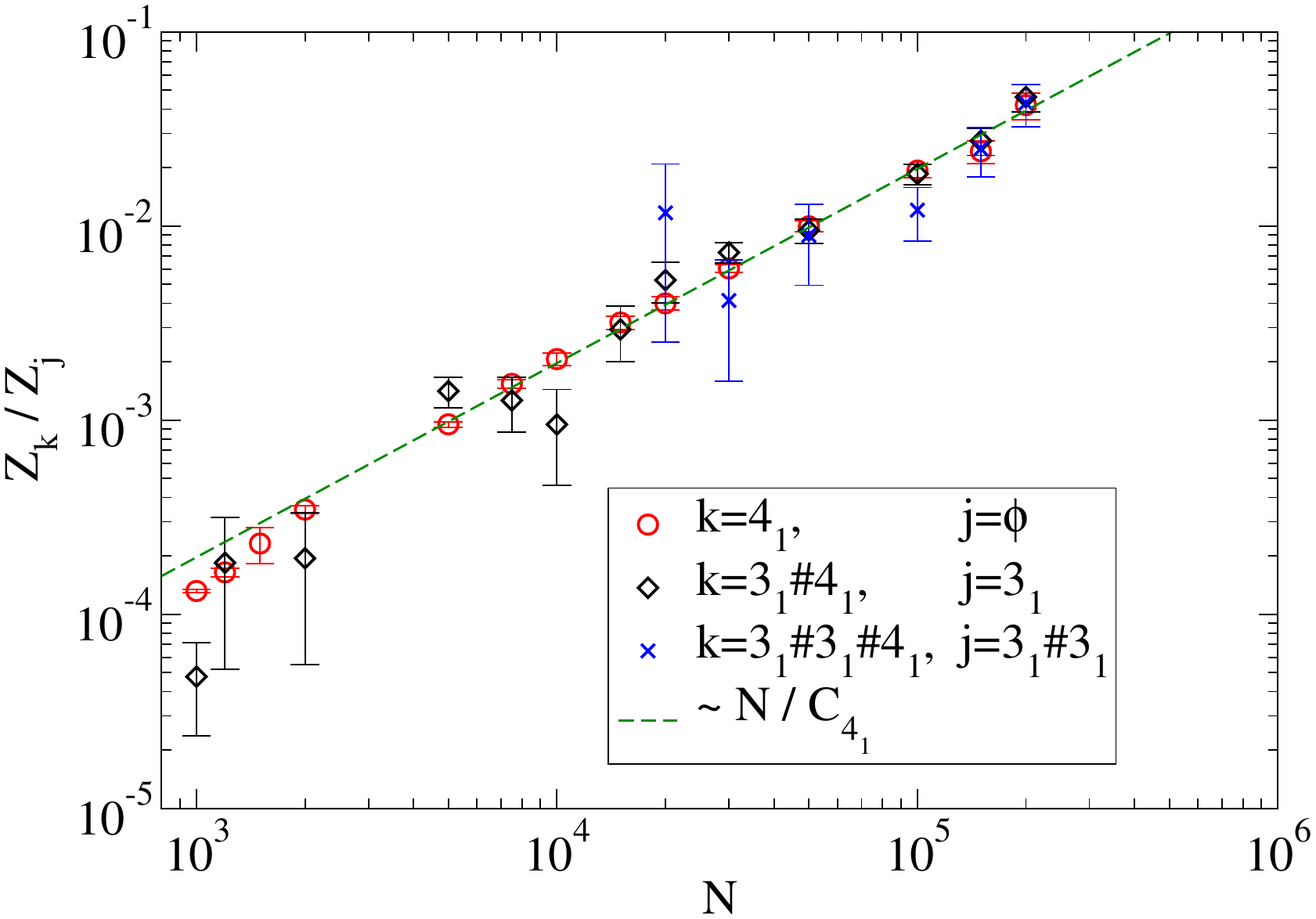}
\vskip 3mm
\caption{
Partition function of knots including a prime component $4_1$ divided
by that of the same knots without the $4_1$, in log-log scale vs $N$.
The straight line is a guide to the eye, scaling $\sim N$.
\label{fig:41}}
\end{center}
\end{figure}
%%%%%%%%%%%%%%%%%%%%%%%%%%%%%%%%%%%%%%%%%%%%%%%%%%%%%%%%%%%%%%%%%%%

The statistics collected for multiple copies of the next simplest knots,
like the $4_1$, is not sufficient to repeat the analysis of equation Eq.~(\ref{eq:31}).
We can however look at Eq.~(\ref{eq:0}) in the case in which, in addition to multiple $3_1$'s, other
prime knots are present. In particular in Fig.~\ref{fig:41} we plot $Z_{4_1}/Z_{\UN}$,  $Z_{3_1\#4_1}/Z_{3_1}$,
and $Z_{3_1\#3_1\#4_1}/Z_{3_1\#3_1}$. As expected, all are consistent with the presence of the term
$N/C_{4_1}$ in the scaling, with $C_{4_1}\simeq 5\times 10^6$.
This result extends to components of
different knot type  the hypothesis of entropic independence between prime components in the statistics
of polygons with a composite knot type $k$.

\subsection{On the connective constant of a class of composite knots}\label{sec:Ck:mu}

So far the only  results available on the limiting entropy of knotted polygons are the rigorous
inequality $\mu_\UN < \mu$ and the conjectured identity $\mu_\UN = \mu_k$, where $\mu_k$ refers to
the connective constant of the subset of polygons having a given knot type $k$. The essential
difference between $\mu$ and $\mu_\UN$ or $\mu_k$ is that in the first case the (infinite) sum over
all topologies is taken into account while for $\mu_\UN$ and $\mu_k$ the topology is kept fixed.

By exploiting Eq.~(\ref{eq:0}) it is tempting to interpolate between the extreme cases $\mu$ and
$\mu_k$ by looking at the statistics of  particular subsets of polygons in which an infinite
(although partial) sum over topologies is considered.

Suppose for example to consider the set of all polygons that can have an arbitrary number of
trefoil components tight in:
\begin{equation}
Z_{{(3_1\#)^{\infty}}}(N) = Z_{3_1}(N) + Z_{3_1\#3_1}(N) + Z_{3_1\#3_1\# 3_1}(N) + \cdots
\label{Z_all_3_1}
\end{equation}
In the limit of large $N$, using Eq. (\ref{eq:0}), we get
\[
Z_{{(3_1\#)^{\infty}}}(N) \simeq \sum_{n=0}^\infty Z_\UN(N) \frac{1}{n_k!}\left(\frac
N{C_{3_1}}\right)^n \simeq Z_\UN(N) e^{N/C_{3_1}}.
\]

By rewriting the exponential factor as $(\mu_\UN e^{1/C_{3_1}})^N = (\mu e^{-1/N_0 + 1/C_{3_1}})^N
= (\mu_{(3_1\#)^{\infty}})^{N}$, since  $C_{3_1}> N_0$, we get $\mu_{(3_1\#)^{\infty}} > \mu_\UN$.
It is interesting to notice that, if we apply the same argument to the set of composite knots made
only by $4_1$ knots, since $C_{4_1} > C_{3_1}
> N_0$, we will get $\mu_{(3_1\#)^{\infty}} > \mu_{(4_1\#)^{\infty}} >\mu_\UN$.
In general we would expect that given two prime knots $k'$ and $k''$ with $C_{k''} > C_{k'}$
\begin{equation}
\mu_{(k'\#)^{\infty}} > \mu_{(k''\#)^{\infty}} > \mu_\UN.
\end{equation}
This can be explained by arguing that each prime knot, being localized, brings the same entropic gain $\sim N$,
but the simplest ones require less entropic cost to be formed.
On the other hand the statistics of topologically unconstrained polygons are, in the large $N$
limit, dominated by extremely complex composite knots made by an arbitrary number of different
prime components. It is then interesting to look at a more complex subsets of polygons whose topology
is characterized by an arbitrary number of $3_1$s and $4_1$s. Clearly
$Z_{(3_1\#)^{\infty},(4_1\#)^{\infty}}(N) > Z_{(3_1\#)^{\infty}}(N)+ Z_{(3_1\#)^{\infty}}(N)$ and
by applying the same argument we obtain $\mu_{(3_1\#)^{\infty},(4_1\#)^{\infty}}= \mu_\UN
e^{1/C_{3_1}+1/C_{4_1}}$. Hence in general we should expect a sequence of the kind
\begin{eqnarray}
\mu_\UN = \mu_{3_1}=\mu_{4_1}=\cdots &<& \cdots < \mu_{(5_1\#)^{\infty}}< \mu_{(4_1\#)^{\infty}}<
\mu_{(3_1\#)^{\infty}}\nonumber
\\  &<&
\mu_{(3_1\#)^{\infty},(4_1\#)^{\infty}}<\mu_{(3_1\#)^{\infty},(4_1\#)^{\infty},(5_1\#)^{\infty}} <
\cdots \nonumber \\ &<& \mu \label{ineq}
\end{eqnarray}

\section{Conclusions}\label{sec:end}

By sampling polygons with $N$ up to $200000$ we have been able to get accurate estimates of the large
$N$  behaviour of the configurational entropy of SAPs with a fixed knot type $k$. We have
corroborated the belief that in good solvent conditions and in the large $N$ limit prime knots are
localized within small regions that slide independently along the unknotted part of the polygon. 
The existence of
each prime component $k$ requires an entropic cost $C_k$ whose dependence on $k$ turns out to be
relatively simple and intriguingly related to the minimal knot length $\ell_k$, i.e the minimal number
of steps necessary to build a knot of type $k$ on the cubic lattice. The above findings allow to
write down a general formula for the partition function of an arbitrary complex composite knots and
to conjecture a sequence of inequalities relating the connective constants of polygons with
different topologies, including families of composite knots. In the future it would be nice to
explore more broadly the asymptotic relation~(\ref{eq:0}) and in particular to test the robustness of
the relation $C_k \propto \exp(\ell_k/3)$ with respect to different polymer models. In particular it
would be interesting to test it in the case of off-lattice polymers  where $\ell_k$ should be
replaced the length of the knot in  its ideal representation
conformations~\cite{Katritch_et_al:1996:Nature,Katritch:1997:Nature:9217153}. Finally we hope that,
inspired by the results presented above, the set of conjectured inequalities in (\ref{ineq}) could
be put on rigorous basis by following new approaches to the problem.

\section*{Acknowledgements:}
We thank A. Grosberg for useful discussions.
M.B. acknowledges financial support from University of Padua
(Progetto di Ateneo n. CPDA083702).

\section*{References}

\bibliography{knot-cost_ref}

\end{document}